\documentclass[AMA,STIX1COL]{WileyNJD-v2}
\usepackage{moreverb}
\usepackage{verbatim}

\articletype{Main paper}%

\received{<22> <January>, <2023>}

\begin{document}
\title{A unified Bayesian framework for interval hypothesis testing in clinical trials}

\author[1]{Abhisek Chakraborty*}
\author[2]{Megan H. Murray}
\author[2]{Ilya Lipkovich}
\author[2]{Yu Du}

\address[1]{\orgname{Department of Statistics, Texas A\&M University}, \orgaddress{\city{College Station}, \state{Texas}, \country{USA}}}
\address[2]{\orgname{Eli Lilly and Company}, \orgaddress{\city{Indianapolis}, \state{Indiana}, \country{USA}}}

\corres{*Abhisek Chakraborty; \email{abhisek\_chakraborty@tamu.edu}}

\abstract[Abstract]{
The American Statistical Association (ASA) statement on statistical significance and P-values \cite{wasserstein2016asa} cautioned statisticians against making scientific decisions solely on the basis of traditional P-values. The statement delineated key issues with P-values, including a lack of transparency, an inability to quantify evidence in support of the null hypothesis, and an inability to measure the size of an effect or the importance of a result. In this article, we demonstrate that the interval null hypothesis framework (instead of the point null hypothesis framework), when used in tandem with Bayes factor-based tests, is instrumental in circumnavigating the key issues of P-values. Further, we note that specifying prior densities for Bayes factors is challenging and has been a reason for criticism of Bayesian hypothesis testing in existing literature. We address this by adapting Bayes factors directly based on common test statistics. We demonstrate, through numerical experiments and real data examples, that the proposed Bayesian interval hypothesis testing procedures can be calibrated to ensure frequentist error control while retaining their inherent interpretability. Finally, we illustrate the improved flexibility and applicability of the proposed methods by providing coherent frameworks for competitive landscape analysis and end-to-end Bayesian hypothesis tests in the context of reporting clinical trial outcomes.
}

\maketitle

\section{Introduction}
The lack of reproducibility of scientific studies has generated increasing apprehension regarding the trustworthiness of assertions about novel discoveries relying on statistical significance \citep{Benjamin2018}. Significant strides have since been made in documenting various factors contributing to this reproducibility crisis. Specifically, the American Statistical Association Statement on Statistical Significance and P-Values \citep{wasserstein2016asa} highlighted that numerous published scientific results rely entirely on P-values, despite the fact that this metric is often misused and misunderstood. As a consequence, several scientific journals have started discouraging the utilization of P-values, and  many statisticians  have even advocated for their abandonment. Informally, a P-value represents the probability, under a specified statistical model, that a statistical summary of the data would be as extreme as or more extreme than the observed value. P-values do not offer a direct measure of support for either the null or alternative hypotheses, The arbitrariness of P-value thresholds for determining statistical significance have been a topic of extensive debate \citep{benjamin2018redefine, johnson2013revised, haaf2019retire, betensky2019pvalue}. In this article, we illustrate that employing a interval null hypothesis framework, rather than the point null hypothesis framework, in conjunction with Bayes factor-based tests, offers an alternative approach for addressing the significant challenges associated with P-values.

Following the talk around the the reproducibility crisis in scientific studies, and American Statistical Association's subsequent statement on P-Values \citep{wasserstein2016asa}, many alternatives to traditional P-value-based testing are already proposed in the literature  both from a frequentist and a Bayesian perspectives, e.g, second-generation P-values \citep{doi:10.1080/00031305.2018.1537893}, e-values \citep{evalues}, d-values \citep{dvalues}, to name a few. Before proceeding further, we present a brief summary of these related ideas. The second-generation P-value (SGPV \citep{doi:10.1080/00031305.2018.1537893}) retains the simplicity of P-values while formally addressing scientific relevance through the utilization of a composite null hypothesis that encompasses null and scientifically trivial effects. Since the majority of spurious findings involve small effects that are technically non-null but practically indistinguishable from the null, the second-generation approach thus significantly diminishes the probability of a false discovery. In simpler terms, SGPVs enhance the reproducibility of scientific outcomes by determining beforehand which potential hypotheses hold practical significance. Additionally, SGPVs offer a more concrete statistical summary of when the data align with null or alternative hypotheses, or when the data render an inconclusive result. E-values \citep{evalues}, on the other hand, represent the outcomes of bets against the null hypothesis. Formally, an e-variable is a non-negative extended random variable with an expected value of at most 1 under the null hypothesis, and an e-value is a specific value taken by such an e-variable. Viewed as a bet against the null hypothesis, the realized value of an e-variable indicates the degree of our success. High success suggests reason to question the truth of the null hypothesis, and the e-value can be seen as the measure of evidence we've gathered against it. The idea of \cite{dvalues} arises from another persistent criticism of the P-values - its susceptibility to being reduced to a very small number with a sufficiently large sample size. This ability to obtain extremely small P-values with increased sample sizes can lead to false positive findings, resulting in conflicting statistically significant reports. Variations in sample sizes also contribute to the lack of reproducibility in scientific studies, where the same studies conducted with different sample sizes yield different P-values. In this regard, \cite{dvalues} proposed using standard deviations instead of standard errors in the computation of the test statistic and tail probabilities to calculate what they term the "D-value."  Owing to SGPV's conceptual and operational simplicity as well as the flexibility of considering interval null hypothesis, it has been widely adopted in reporting clinical trial outcomes.

Importantly, second-generation P-values have significant predecessors in the literature that offer the flexibility of considering interval null hypotheses, such as equivalence tests \citep{schuirmann1987two, MEYNERS2012231}. In these tests, the null hypotheses are defined in terms of effect sizes that are considered sufficiently interesting, as indicated by an equivalence bound. One straightforward approach to equivalence testing is the Two One-Sided Tests (TOST) procedure \citep{schuirmann1987two}. In the TOST procedure, upper $\delta_{u}$ and lower $\delta_{l}$ equivalence bounds are specified based on the smallest effect size of interest. Two composite null hypotheses are tested: $H_{01}: \delta\leq -\delta_{1}$ and $H_{02}: \delta\geq \delta_{2}$. When both these one-sided tests can be statistically rejected, we conclude that $-\delta_{1} < \delta < \delta_{2}$, or the observed effect is within the equivalence bounds and is statistically smaller than any effect that is considered meaningful and practically equivalent. In the clinical setting, non-inferiority tests, a widely adopted form of equivalence testing \citep{riechelmann2019noninferiority, cuzick2022interpreting}, are commonly employed. These tests are utilized when introducing a new drug that is posited to be no worse than currently the best available option in the market in terms of efficacy, but may offer additional advantages, such as an improved safety profile.

On the Bayesian side, Bayes factors \citep{kass1995bayes, jeffreys1961theory, morey2011bayes, rouder2009bayesian, wagenmakers2010bayesian, johnson2023bayes, chakraborty2024differentially} provide a valuable alternative to P-values for expressing the results of hypothesis tests. They directly quantify the relative evidence in favor of competing hypotheses, effectively addressing the persistent criticisms of P-values. Additionally, Bayes factors offer the flexibility to consider multiple competing interval hypotheses. Bayes factors represent the ratio of the marginal probability assigned to data by competing hypotheses. When combined with prior odds assigned between hypotheses, they yield an estimate of the posterior odds for the validity of each hypothesis. However, selecting a prior for the parameter of interest under various competing hypotheses can be challenging. As a result, several Bayes factors have been introduced based on ``default" alternative prior densities. Nevertheless, the value of a Bayes factor depends on the particular alternative prior density chosen for its calculation, and justifying or interpreting a single default choice can be problematic. Furthermore, the numerical computation of Bayes factors can be computationally demanding. To that end, a series of articles \citep{723ff713-f247-314c-9ec9-911ba102f213, 937ddeeb-bd63-3520-be66-7bc978d22bbd, johnson2023bayes} introduced substantial modifications aimed at improving the reporting of scientific findings by defining Bayes factors directly from standard test statistics. Under the point null hypotheses, the distribution of these test statistics is known. Under alternative hypotheses, the asymptotic distributions of these test statistics depend only on scalar-valued non-centrality parameters. As a result, the specification of the prior density that characterizes the alternative hypothesis is simplified. The prior densities suggested for non-centrality parameters are specific instances of non-local alternative prior densities \citep{johnson2023bayes, chakraborty2023bayesian}. These particular densities have a unique property that they are zero when the non-centrality parameter is zero. This characteristic enables a faster accumulation of evidence in support of both genuine null and true alternative hypotheses \citep{efficientalternative}. We borrow from the existing literature on Bayesian testing and propose Bayes factors based on test statistics \cite{723ff713-f247-314c-9ec9-911ba102f213, 937ddeeb-bd63-3520-be66-7bc978d22bbd} for the interval null hypothesis testing setup and adopt it to accomplish several critical tasks with regard to reporting and interpretation of clinical trial outcomes.

Section \ref{2way_testing} introduces our proposed methodology in a common two-way hypothesis testing setup, discusses the prior specification, and associated automated hyper-parameter tuning schemes. Section \ref{3way_testing} extends the methodology to the multiple competing hypothesis testing setup and presents an application in testing for the superiority, equivalence, and inferiority of an investigational diabetes treatment, Basal Insulin Peglispro (BIL), compared to Insulin Glargine in patients with Type II Diabetes Mellitus. In section \ref{ssec:metanalysis}, we extend the proposed methodology to the meta-analysis setup and discuss several applications in clinical trials, including end-to-end Bayesian hypothesis testing and competitive landscape analysis, along with real data analysis. We conclude with a discussion.

\section{Bayes Factor (BF) based on Test Statistics}\label{2way_testing}

For the purpose of simplicity, we present the proposed methodology within the framework of a one-sample setup. Later, we elaborate on our proposals, extending them to more practical two-sample setups using real data examples. Importantly, our approach involves defining Bayes factors directly based on the common t-statistic. Generalizing the following stylized one-sample example to accommodate various modifications simply requires calculating the degrees of freedom and non-centrality parameter of the t statistic under the new setup.

To that end, suppose we observe data $X_1,\ldots, X_n\sim \mbox{N}(\mu, \sigma^2)$. For a fixed $\delta\in\mathrm{R}^{+}$, we want test the null and the alternative  hypotheses of the form
\begin{align}
    \mbox{H}_0: \mu\in(-\delta, \delta)
    \quad \text{against}\quad 
    \mbox{H}_1: \mu\in (-\infty, -\delta]\cup [\delta, \infty).
\end{align}
Bayesian testing procedures involve elicitation of the prior on (i) the hypotheses of interest, denoted by $\mbox{P}(\mbox{H}_i)$, and (ii) the parameters $(\mu, \sigma^2)$ of interest, separately under null and the alternative hypotheses, denoted by $\Pi_{\mbox{H}_0}$ and $\Pi_{\mbox{H}_1}$ respectively. Then, the marginal likelihood under $\mbox{H}_i$ is calculated via
\begin{equation}
m_i(x_{1:n}) = \int_{\mu, \sigma^2} \bigg[\prod_{i=1}^n\mbox{N}(x_i\mid\mu, \sigma^2)\times \Pi_{\mbox{H}_i}(\mu, \sigma^2)\bigg]d\mu\ d\sigma^2,\ i= 0,1,
\end{equation}
and the posterior probability assigned to the hypothesis $\mbox{H}_i$ is calculated via
\begin{equation}
\mbox{P}(\mbox{H}_i\mid x_{1:n}) = \frac{m_i(x_{1:n})\times \mbox{P}(\mbox{H}_i)}{m_0(x_{1:n})\times \mbox{P}(\mbox{H}_0) + m_1(x_{1:n})\times \mbox{P}(\mbox{H}_1)}, \ i= 0,1.
\end{equation}
We often assume $\mbox{P}(\mbox{H}_i) = 0.5$, a-priori. In practice, we have the flexibility to modify $\mbox{P}(\mbox{H}_i)$ on the basis of accumulated evidence for certain class of drugs. 

Importantly, the specification of the prior on the parameter of interest under various competing hypotheses is subjective, and it is generally difficult to justify or interpret any default choice. In addition, the numerical calculation of Bayes factors can often be challenging. The article \cite{johnson2023bayes} proposed several modifications to the existing Bayes factor methodology designed to enhance the reporting of scientific findings. In particular, \cite{johnson2023bayes} defines Bayes factors directly based on standard test statistics. Under point null hypotheses, the distribution of these test statistics is known. Under alternative hypotheses, the asymptotic distributions of these test statistics depend only on scalar-valued non-centrality parameters. Thus, the specification of the prior density that defines the alternative hypothesis is simplified. For example, suppose we want to test the point null hypothesis $\mbox{H}_0: \mu = 0$ versus the alternative $\mbox{H}_1: \mu \neq 0$. Then the common test statistic $t \sim \mbox{t}(\nu, 0)$ under $\mbox{H}_0$; and $t \sim \mbox{t}(\nu, \lambda)$ under $\mbox{H}_1$, where $\nu = n - 1$ and $\lambda$ are the degrees of freedom and the non-centrality parameter of the student's $t$ distribution, respectively. Johnson et al. \cite{johnson2023bayes} proposes to construct Bayes factors directly based on $t$ statistics via putting a prior directly on the non-centrality parameter $\lambda$.

Let us return to the intended exercise in the article, e.g., testing the interval null hypothesis $\mbox{H}_0: \mu\in(-\delta, \delta)$ against $\mbox{H}_1: \mu\in (-\infty, -\delta]\cup [\delta, \infty)$. Under this setup, although the $t$ statistic still follows $t \sim \mbox{t}(\nu, \lambda)$ with degrees of freedom $\nu = n - 1$, the non-centrality parameter $\lambda$ is non-zero both under $\mbox{H}_0$ and $\mbox{H}_1$. In this article, to develop a Bayes factor-based testing framework, we propose to put different priors on $\lambda$ under $\mbox{H}_0$ and $\mbox{H}_1$. Details of the prior specification are given in the following.

\subsection{Non-local Priors and Hyper-parameter Tuning}\label{prior_basics}
\begin{figure}[!htb]
        \center{\includegraphics[height = 7cm, width = 12cm]
        {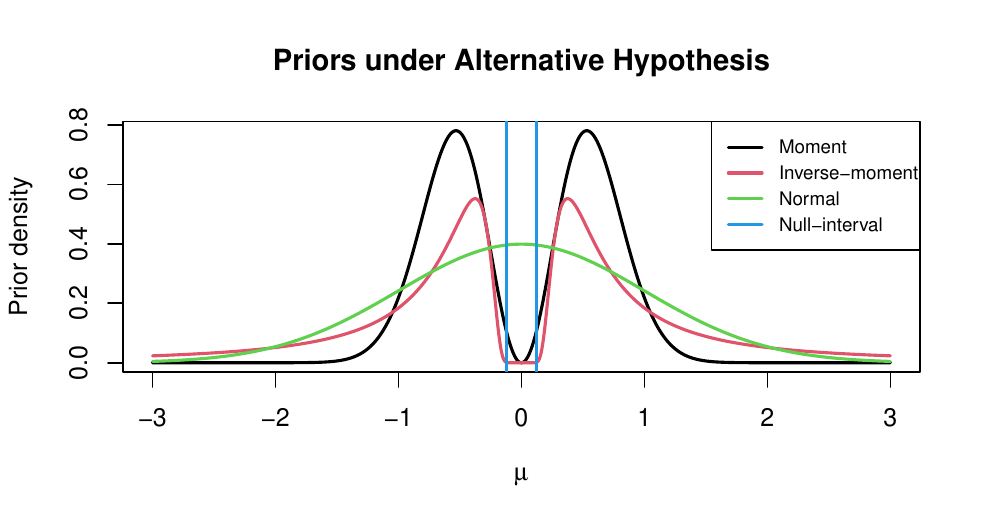}}
        \caption{\label{fig:priors} \textbf{Non-local priors versus Local Priors.} Non-local densities (e.g moment, inverse moment densities), unlike local priors (e.g normal, t, Laplace), put little to no prior mass to values of non-centrality parameters that are consistent with the null hypothesis.}
\end{figure}

Under the alternative hypothesis, we may be inclined to put local priors (e.g., Normal, Laplace, Student's-$t$, etc.) on the non-centrality parameter $\lambda$. Although such local priors are routinely utilized in Bayesian testing literature, this is subject to criticism if used in our setup, since these priors put substantial mass on values of non-centrality parameters that are consistent with the null hypothesis. On the contrary, if non-local densities (e.g., moment, inverse moment densities) are used as priors on the non-centrality parameter $\lambda$ under the alternative hypothesis, they put little to no prior mass on values of non-centrality parameters that are consistent with the null hypothesis. In what follows, we denote the non-local prior densities via $\Pi_{\mbox{H}_i}(\mu\mid \tau)$, where $\tau$ is a prior hyper-parameter. Refer to Figure \ref{fig:priors} for the visual representation of the local and non-local priors.

In this article, we shall focus on two specific choices of the non-local priors: (i) the normal moment prior with probability density function
\begin{align}
    J(u\mid \tau^2)\ = \ \frac{u^2}{\sqrt{2\pi}\tau^3}\ \exp\bigg(-\frac{u^2}{2\tau^2}\bigg), \ u\in\mathbf{R},
\end{align}
or the (ii) the normal inverse moment prior with probability density function
\begin{align}
    J(u\mid \tau^2)\ = \ \frac{\tau^{1/2}}{\Gamma\big(\frac{1}{2}\big)}\ \ u^{-2}\exp\bigg(-\frac{\tau}{u^2}\bigg), \ u\in\mathbf{R},
\end{align}
where  $\tau>0$ is a prior hyper-parameter. Before we move on, we describe an objective prior hyper-parameter tuning scheme next.

\begin{figure}[h]
  \includegraphics[width=1\columnwidth, height = 5cm]
    {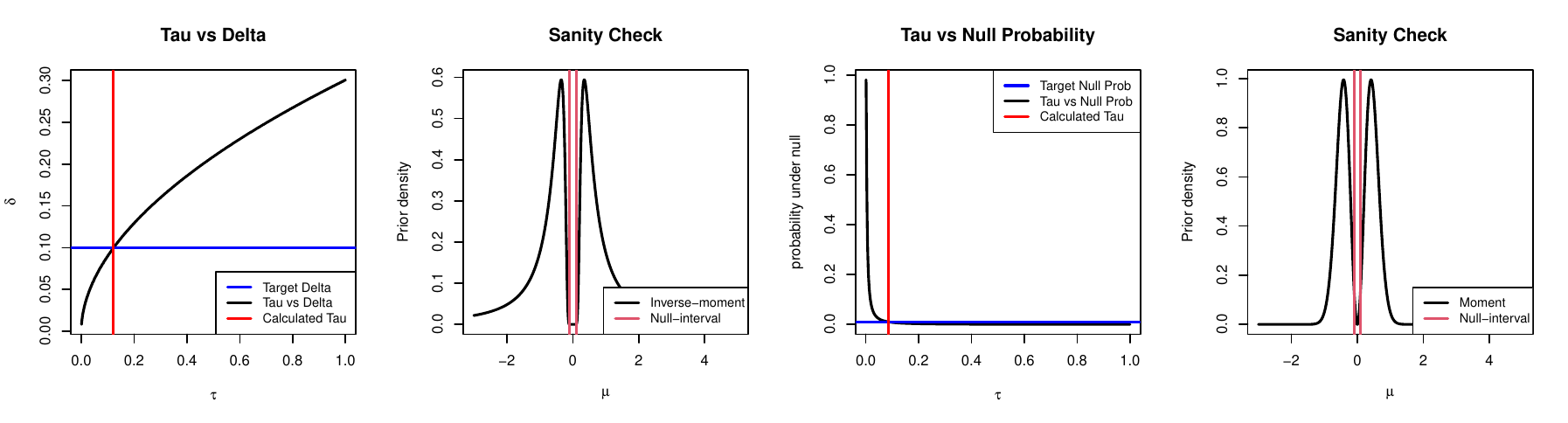}\hfill

  \caption{\textbf{Hyper-parameter tuning.}
  \textbf{(1st \& 2nd plot, Moment prior)}. The figure demonstrates the  hyper-parameter tuning exercise for moment priors, where we fixed $\delta = 0.1$ and $\varepsilon = 0.01$. 
  \textbf{(3rd \& 4th plot , Inverse moment prior)}. We present the hyper-parameter tuning scheme for the inverse moment prior, where we fixed $\delta = 0.1$. We have developed automated sub-routines to choose the hyper-parameter $\tau$, given a target value of $\delta$.}\label{fig:hyperparameter}
\end{figure}
Note that the moment prior does put some probability into the null interval $(-\delta, \delta)$, and we denote this probability by $\Pi_{\mbox{H}_1}(\mu\in(-\delta, \delta)\mid \tau)$. But for a fixed $\delta$, we can choose the hyperparameter $\tau$ such that $\Pi_{\mbox{H}_1}(\mu\in(-\delta, \delta)\mid \tau)=\varepsilon$, where $\varepsilon$ is a small fixed target value. The figure \ref{fig:hyperparameter} demonstrates this exercise for $\delta = 0.1$ and $\varepsilon = 0.01$. We have developed automated subroutines to choose the hyperparameter $\tau$, given a set of target values of $(\delta, \varepsilon)$. As long as $\varepsilon$ is not "too" big, the moment prior does not put much mass into the null interval $(-\delta, \delta)$. On the other hand, the inverse-moment prior puts absolutely no mass within a symmetric interval around $0$. For a fixed $\delta$, we can choose the prior hyperparameter $\tau$ such that $\Pi_{\mbox{H}_1}(\mu\in(-\delta, \delta)\mid \tau)=0$. We now have all the ingredients required to showcase our proposal in action and compare it with competing methods available in the literature via elaborate simulation studies.

\subsection{Numerical Experiment 1} 
\begin{figure}[!htb]
        \center{\includegraphics[height = 5cm, width = 15cm]
        {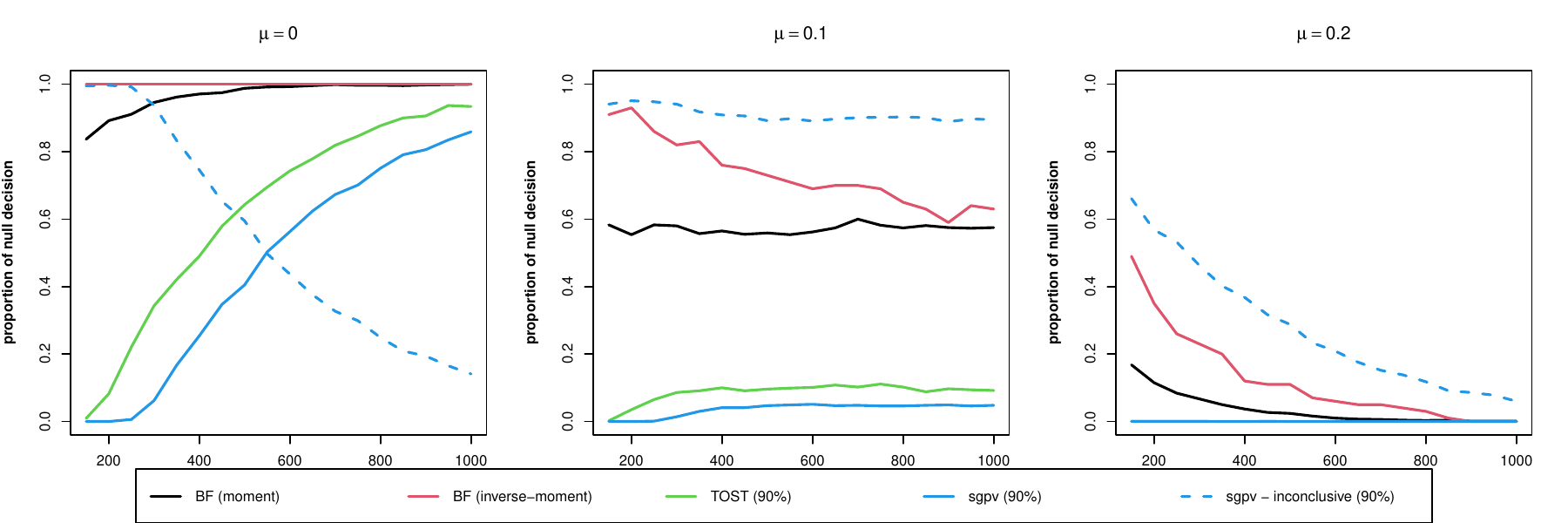}}
        \caption{\label{consitency_95} We present the proportion of times we conclude that the null is true in repeated simulations versus the sample size, for varying simulation set ups, e.g, the data is generated with $\mu\in\{0.0, 0.1, 0.2\}$. When null is true (i.e $\mu= 0.0$), we want  this proportion to converge to $1$ as $n\to\infty$. When alternative is true (i.e $\mu= 0.2$), we want this proportion to converge to $0$ as $n\to\infty$. For BF, if $\mbox{P}(\mbox{H}_0\mid x_{1:n})> 0.5$, we conclude null.}\label{two_way_50}
\end{figure}

\begin{figure}[!htb]
        \center{\includegraphics[height = 5cm, width = 15cm]
        {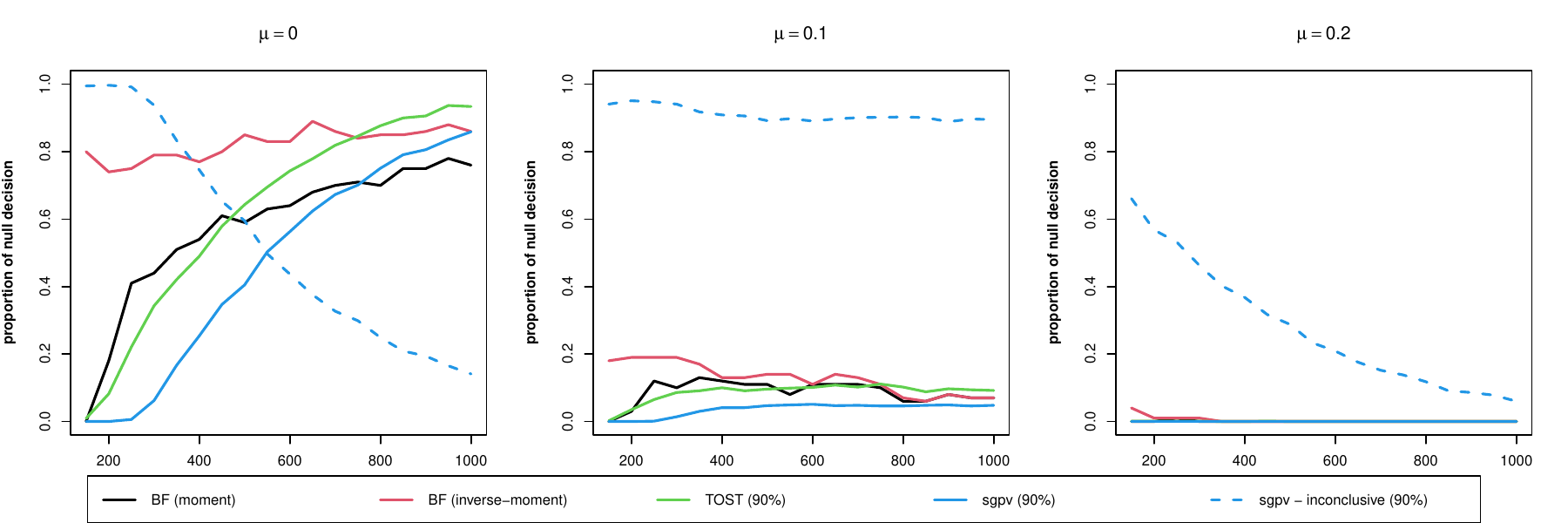}}
        \caption{\label{consitency_95} We present the proportion of times we conclude that the null is true in repeated simulations versus the sample size, for varying simulation set ups, e.g, the data is generated with $\mu\in\{0.0, 0.1, 0.2\}$. 
         When null is true (i.e $\mu=0.0$), we want  this proportion to converge to $1$ as $n\to\infty$. When alternative is true (i.e $\mu= 0.2$), we want this proportion to converge to $0$ as $n\to\infty$. For BF, if $\mbox{P}(\mbox{H}_0\mid x_{1:n})> 0.7$, we conclude null.}\label{two_way_70}
\end{figure}

In this section, under varying data generation schemes, we present a comparison of the proposed Bayes factor-based tests with moment and inverse moment priors equipped with the hyperparameter tuning scheme introduced in subsection \ref{prior_basics}, with two frequentist approaches, e.g., Second-generation P-values, two one-sided t-test. We generate data $X_1,\ldots, X_n\sim \mbox{N}(\mu, 1)$ such that the mean $\mu\in\{0.0, 0.02, 0.1, 0.2\}$ and sample size $n\in\{150, 200, 250, \ldots, 1000\}$. We want to test the hypothesis $\mbox{H}_0: \mu\in(-\delta, \delta)$ against $\mbox{H}_1: \mu\in (-\infty, -\delta]\cup [\delta, \infty)$ where $\delta = 0.1$. So, in practice, $\mu\in\{0.0, 0.02\}$ corresponds to the case when the null is true, $\mu = 0.1$ corresponds to the borderline case, and finally, $\mu = 0.2$ corresponds to the case when the alternative is true. The level of significance for the frequentist testing methods (second-generation P-values, TOST) is set at $\alpha$. For BF, if $\mbox{P}(\mbox{H}_0\mid x_{1:n})> \kappa$, we conclude null. Refer to Figures \ref{two_way_50} and \ref{two_way_70} for results, and discussion for $\alpha = 0.1$ and $\kappa\in\{0.5, 0.7\}$. The simulations confirm that Bayesian tests based on BF can be calibrated via an adaptive choice of $\kappa$, to ensure frequentist error controls.

\section{Three (Multiple) competing hypothesis}\label{3way_testing}
In many practical applications, we need to evaluate several (more than two) competing hypotheses in light of the observed data. For example, in clinical trials, we may want to test for inferiority, equivalence, and superiority of the administered drug compared to an existing drug. No natural extension of the frequentist testing procedures is available in the literature to tackle the problem. However, we can seamlessly extend our methods to scenarios with more than two hypotheses to ensure a coherent clinical evaluation of the hypotheses of interest.

To that end, suppose we observe data $X_1,\ldots, X_n\sim \mbox{N}(\mu, \sigma^2)$. For a fixed $\delta\in\mathrm{R}^{+}$, we want test the  hypotheses of the form
\begin{align}
    \mbox{H}_1: \mu\in(-\infty,\ \delta_1),
    \quad 
    \mbox{H}_2: \mu\in (\delta_1, \delta_2],
    \quad
    \mbox{H}_3: \mu\in (\delta_2,\ \infty].
\end{align}
Bayesian testing procedures involve elicitation of the prior on (i) the hypotheses of interest, denoted by $\mbox{P}(\mbox{H}_i), \ i = 1, 2, 3$, and (ii) the parameters $(\mu, \sigma^2)$, separately under the different competing hypotheses, denoted by $\Pi_{\mbox{H}_i}, \ i = 1, 2, 3$. Then, the marginal likelihood under $\mbox{H}_i$ is calculated via
\begin{equation}
m_i(x_{1:n}) = \int_{\mu, \sigma^2} \bigg[\prod_{i=1}^n\mbox{N}(x_i\mid\mu, \sigma^2)\times \Pi_{\mbox{H}_i}(\mu, \sigma^2)\bigg]d\mu\ d\sigma^2,\ i= 1, 2, 3,
\end{equation}
and the posterior probability assigned to the hypothesis $\mbox{H}_i$ is calculated via
\begin{equation}
\mbox{P}(\mbox{H}_i\mid x_{1:n}) = \frac{m_i(x_{1:n})\times \mbox{P}(\mbox{H}_i)}{\sum_{i=1}^3 m_i(x_{1:n})\times \mbox{P}(\mbox{H}_i)}, \ i= 1, 2, 3.
\end{equation}
We often assume $\mbox{P}(\mbox{H}_i) = 1/3$ a priori. As earlier, we have the flexibility to modify $\mbox{P}(\mbox{H}_i)$ based on accumulated evidence for certain kinds of drugs. Further, owing to issues discussed in Section \ref{2way_testing}, we construct Bayes factors directly based on the $t$-test statistic. The $t$ statistic still follows $t \sim \mbox{t}(\nu, \lambda)$ with degrees of freedom $\nu = n - 1$, and the non-centrality parameter $\lambda$ is non-zero both under $\mbox{H}_0$ and $\mbox{H}_1$. Like earlier, to develop a Bayes factor-based testing framework, we propose to put different priors on $\lambda$ under $\mbox{H}_i, \ i = 1, 2, 3$. Half non-local densities (moment, inverse moment densities), instead of non-local prior densities utilized in Section \ref{2way_testing}, are used as priors under the hypotheses $\mbox{H}_1, \mbox{H}_3$, to ensure that little to no prior mass is assigned to values of non-centrality parameters that are consistent with the hypothesis $\mbox{H}_2$. Since $\delta_1$ may not be equal to $\delta_2$, we slightly modify the hyperparameter tuning schemes described in Section \ref{prior_basics}.

\subsection{Numerical Experiment 2}

\begin{figure}[!htb]
        \center{\includegraphics[height = 5cm, width = 14cm]
        {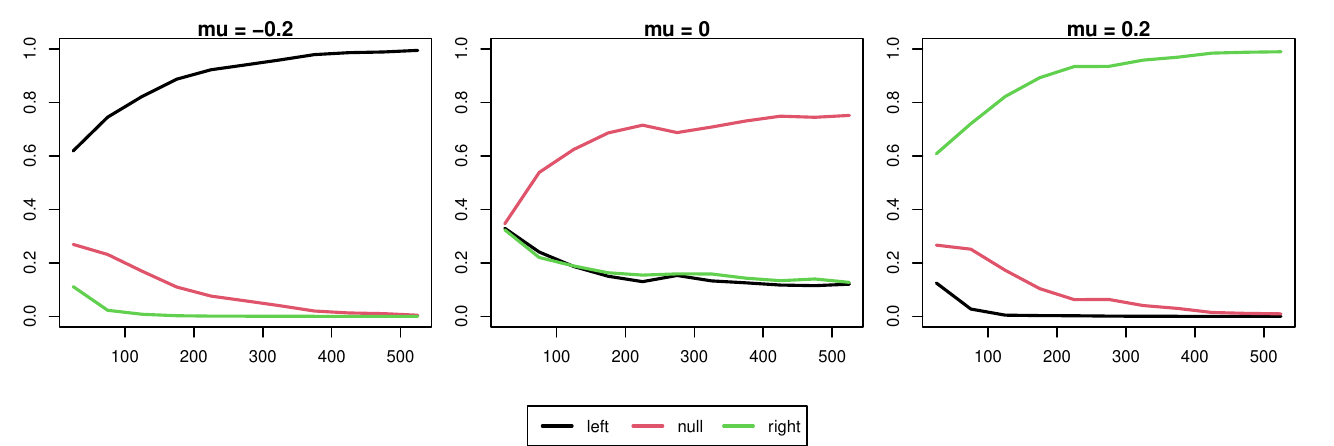}
        }
        \caption{\label{3way_mom}\textbf{Simulation for 3-way testing (moment prior)}. We present the posterior probability of the three competing hypotheses, in varying simulation set ups, i.e,  when the data is generated with varying $\mu\in\{-0.2,  0.0,  0.2\}$.}
\end{figure}

\begin{figure}[!htb]
        \center{\includegraphics[height = 5cm, width = 14cm]
        {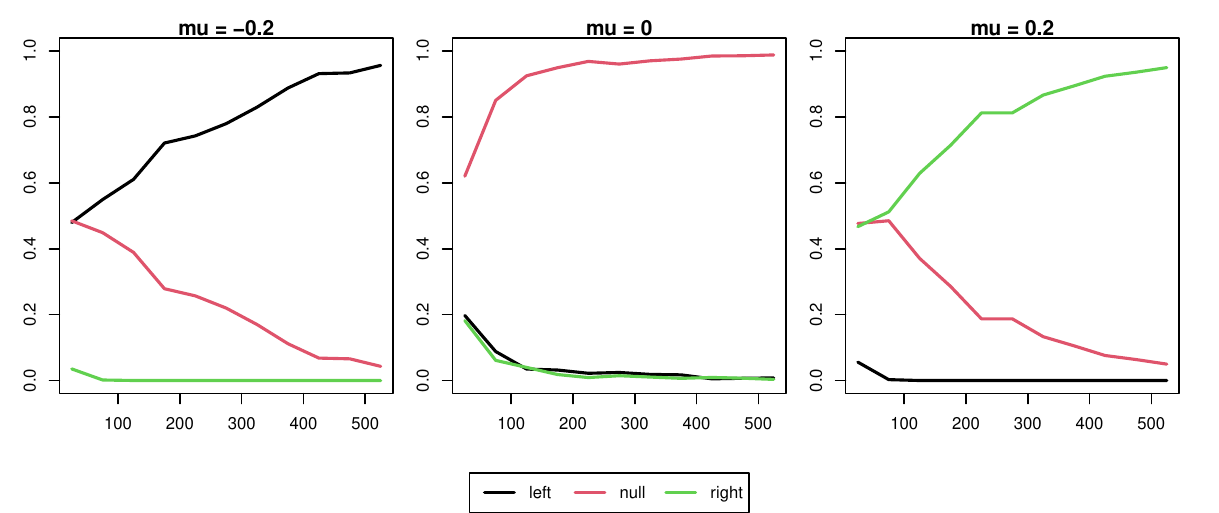}}
        \caption{\label{3way_imom}\textbf{Simulation for 3-way testing (inverse moment prior)}. We present the posterior probability of the three competing hypotheses, in varying simulation set ups, i.e,  when the data is generated with varying $\mu\in\{-0.2,  0.0,  0.2\}$ }
\end{figure}

We present a comparison of the Bayes factor based methods with moment and inverse moment priors, under varying data generation schemes. We generate data $X_1,\ldots, X_n\sim \mbox{N}(\mu, 1)$ such that $\mu\in\{-0.2, 0.0,  0.2\}$ and $n\in\{25, 75, 125, \ldots, 525\}$, and  we want to test the competing hypotheses $\mbox{H}_1 (\text{left}): \mu\leq \delta_1, \mbox{H}_2 (\text{null / middle}): \delta_1<\mu< \delta_2, \mbox{H}_3 (\text{right}): \mu\geq \delta_2$,  where $\delta_1 = -0.1,\ \delta_2 = 0.1$. So, in practice, (i) $\mu  = -0.2$ correspond to the  case when $\mbox{H}_1$ is true,  (ii) $\mu= 0.0$ correspond to the case when $\mbox{H}_2$ is true, and finally (iii)  $\mu = 0.2$ correspond to the case when $\mbox{H}_3$ is true. If $\mbox{P}(\mbox{H}_i\mid x_{1:n})> \max_{j\neq i}[\mbox{P}(\mbox{H}_j\mid x_{1:n})]$, we conclude $H_i$ is true. Refer to Figures \ref{3way_mom} and \ref{3way_imom} that show that the highest posterior probability is assigned to the correct hypotheses, both for the moment and inverse moment prior based approaches. Use of inverse moment prior, compared to the moment prior, seems facilitate quicker accumulation of evidence in favour of true equivalence.

\subsection{Application: Evaluation of the IMAGINE 5 (BIL, Phase-3) study}\label{ssec:bidj}

Study IMAGINE 5 was a Phase 3, open-label, multicenter, multinational, randomized, controlled, parallel design trial with the aim of comparing an investigational diabetes insulin treatment, Basal Insulin Peglispro (BIL), to Insulin Glargine in basal-only insulin-treated patients with Type-II Diabetes Mellitus who are treated with 0 to 3 OAMs \citep{bidjimagine52018}. In this study, BIL and Insulin Glargine were used alone or in combination with pre-study OAM(s). Sulfonylureas (SU) and meglitinides, dipeptidyl peptidase-4 (DPP-IV) inhibitors, biguanides, and alpha-glucosidase inhibitors were continued in combination with insulin at pre-study doses. Pioglitazone was continued in accordance with local regulation. Patients taking rosiglitazone were excluded from the trial. Approximately 426 patients were randomized into this study. Patients were required to complete a screening visit (Visit 1) and a pre-randomization visit (Visit 2) prior to randomization in the trial. At Visit 3, approximately 426 eligible patients were randomized in a 2:1 ratio (BIL:Insulin Glargine) to one of the two treatment arms in this study for 52 weeks of treatment. Patients were administered their first study insulin dose at bedtime on the night of Visit 3; thereafter, patients were administered the study insulin dose at bedtime at approximately the same time every night. Patients who were transitioning from pre-study morning-only administration of basal insulin administered $50\%$ of their first daily dose of study basal insulin at the investigative site on the morning of randomization and the remaining $50\%$ of the dose that evening. For both study insulins, the insulin doses were increased according to an algorithm \citep{riddle2003treat} during the first 26 weeks of the study and thereafter according to investigator judgment. During the course of the study, a Data Monitoring Committee (DMC) comprised of individuals external to Eli Lilly and Company (Lilly) monitored, in an unblinded manner, the safety of BIL and recommended changes to the protocol, including termination, when necessary.

\begin{figure}[!htb]
        \center{\includegraphics[height = 6cm, width = 15cm]
        {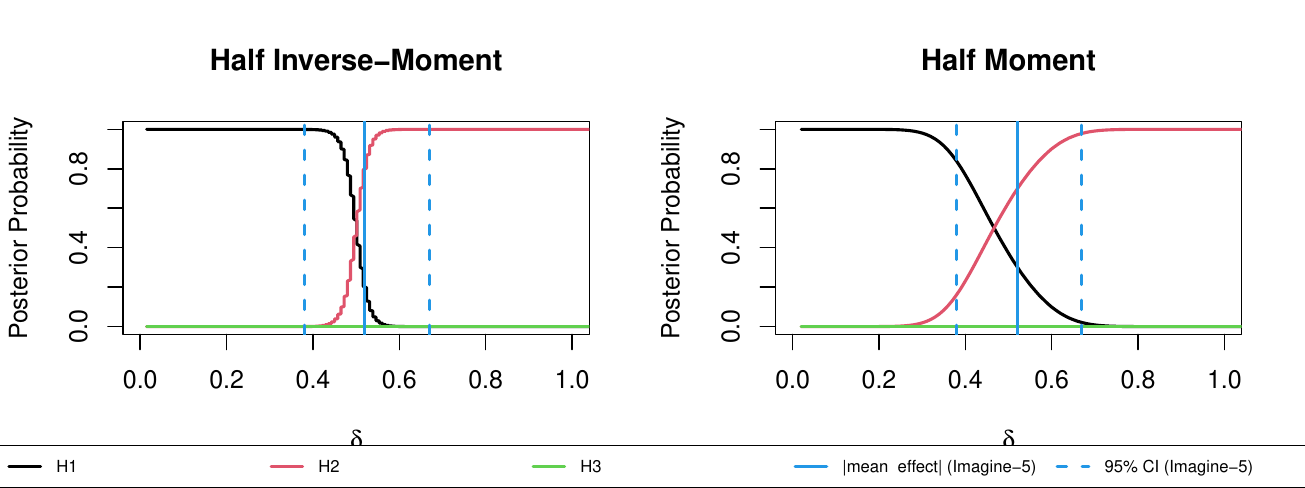}}
        \caption{\label{BIDJ_moment}\textbf{IMAGINE 5 study (moment and inverse moment prior).}  We present the posterior probability of the three competing hypotheses, e.g, superiority ($H_1$), equivalence ($H_2$), inferiority ($H_3$) of BIL compared to Insulin Glargine. }
\end{figure}

The primary objective of this study was to demonstrate that BIL was noninferior to Insulin Glargine for the change in hemoglobin A1c (HbA1c) from baseline to 26 weeks of treatment in patients with T2DM treated with basal insulin alone or in combination with OAM(s) using a noninferiority margin (NIM) of $0.4\%$. To extend the analysis for superiority, equivalence, and inferiority of BIL compared to Insulin Glargine, we can test the competing hypotheses $\mbox{H}_1: \mu_1 -\mu_2\leq \delta_1 = -0.4,\ \mbox{H}_2: -0.4 = \delta_1<\mu_1 - \mu_2< \delta_2 = 0.4,\ \mbox{H}_3: \mu_1 - \mu_2\geq \delta_2 = 0.4$ respectively, where $\mu_1$ and $\mu_2$ refer to the change in hemoglobin A1c (HbA1c) from baseline to 26 weeks of treatment in patients with T2DM, with the administration of BIL and Insulin Glargine, respectively. We utilize Bayes factor based on the two-sample $t$-statistic that follows $t(\nu = n_1 + n_2 -2,\ \lambda)$. Here, $n_1$ and $n_2$ are the sample sizes corresponding to the two groups; and $\nu$ and $\lambda$ are the degrees of freedom and non-centrality parameter of the Student's $t$ distribution respectively. We put half non-local priors on $\lambda$ under $H_1$ and $H_3$, and a flat prior under $H_2$. If $\mbox{P}(\mbox{H}_i\mid x_{1:n})> \max_{j\neq i}[\mbox{P}(\mbox{H}_j\mid x_{1:n})]$, we conclude $H_i$ is true. Refer to Figure \ref{BIDJ_moment} that shows the posterior probability assigned to the three competing hypotheses for varying values of $\delta := |\delta_1| = |\delta_2|$, both for the moment and inverse moment prior based approaches. At $\delta = 0.4\%$, both approaches provide the maximum posterior probability to the hypothesis that BIL is superior to Insulin Glargine with respect to the change in hemoglobin A1c (HbA1c) from baseline at 26 weeks of treatment in patients with T2DM.

\section{Meta analysis}\label{ssec:metanalysis}
In many practical applications, it is common to combine information from several replicated studies \citep{borenstein2009introduction, thompson1999explaining}. Unlike P-value-based approaches, Bayes factor-based methods can be seamlessly used to integrate evidence in favor of different competing hypotheses across studies. Suppose we have $r$ independent replicated studies. Based on the replications, we want to test the $m$ competing hypotheses:
\begin{align}
   \mbox{H}_t: \mu\in\Theta_t \quad t = 1,\ldots, m; \quad \text{such that}\quad  \bigcup_{t=1}^m \Theta_t = \mathbf{R}.
\end{align}
Since the replicated studies are independent, the pooled marginal likelihood of a hypothesis in the meta-analysis would correspond to the product of the marginal likelihood of the hypothesis in the $r$ individual studies. This makes the integration of the evidence in favor of the competing hypotheses from replicated studies exceptionally simple. That is, for the $k$-th replicate, the marginal likelihood of $\mbox{H}_t$ is calculated via:
\begin{equation}
m^{(k)}_t\big(x^{(k)}_{1:n}\big) = \int_{\mu_{(k)}, \sigma_{(k)}^2 } \bigg[\prod_{i=1}^n\mbox{N}(x^{(k)}_i\mid\mu_{(k)}, \sigma_{(k)}^2)\bigg] \Pi_{\mbox{H}_t}(\mu_{(k)},\ \sigma_{(k)}^2)\ d\mu_{(k)}\ d \sigma_{(k)}^2;\quad t= 1,\ldots, m;\ k = 1,\ldots, r.
\end{equation}
For the meta-data, the marginal likelihood of $\mbox{H}_t$ is calculated via $\prod_{k=1}^r m^{(k)}_t\big(x^{(k)}_{1:n}\big)$; and the posterior probability assigned to the hypothesis $\mbox{H}_t$ is calculated by:
\begin{equation}\label{eqn:meta_analysis}
\mbox{P}(\mbox{H}_t\mid \{x^{(k)}_{1:n}\}_{k=1}^r) = \frac{\mbox{P}(\mbox{H}_t)
\times\prod_{k=1}^r  m^{(k)}_t\big(x^{(k)}_{1:n}\big)}{\sum_{t^{\prime} = 1}^m\ \mbox{P}(\mbox{H}_{t^{\prime}})
\times \prod_{k=1}^r m^{(k)}_{t^{\prime}}\big(x^{(k)}_{1:n}\big)}, 
\end{equation}
where $\mbox{P}(\mbox{H}_{t^{\prime}})$ denotes the prior on $\mbox{H}_t,\ t=1, \ldots, m$.

\subsection{Numerical Experiment 3}
\begin{figure}[h]
  \includegraphics[width=1\columnwidth, height = 8cm]
    {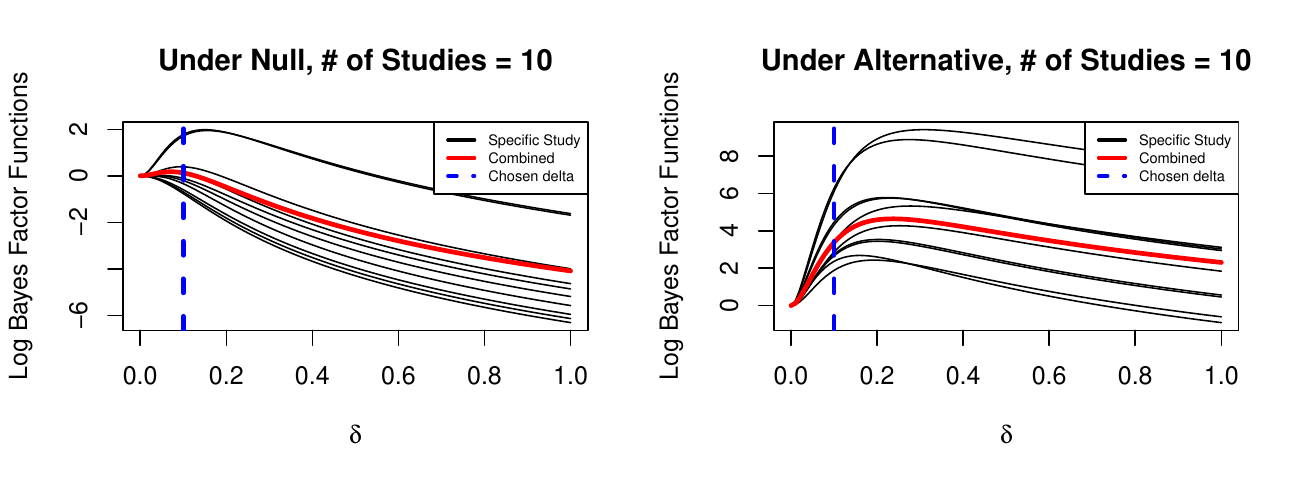}\hfill
  \caption{\label{plot:meta_sim_mom}\textbf{Meta analysis (Moment prior).} \textbf{Left Panel.} The means $\mu_k,\ k =1,\ldots, 10$ are generated at random from $\mbox{Uniform}(0.0, 0.1)$. For varying $\delta$, we present the log Bayes factor for the individual studies,  as well as the meta analysis. Further, if we fix $\delta = 0.1$, the above data generating scheme can be termed as "data generated under null". The combined Bayes factor function (in red) attains its maxima within the null interval (to the left of the dotted blue line), as expected. \textbf{Right panel.} The means $\mu_k,\ k =1,\ldots, 10$ are generated at random from $\mbox{Uniform}(0.2, 0.3)$. For varying $\delta$, we present the log Bayes factor for the individual studies,  as well as the meta analysis. Further, if we fix $\delta = 0.1$, the above data generating scheme can be termed as ``data generated under alternative''. The combined Bayes factor function (in red) attains its maxima outside the null interval (to the right of the dotted blue line), as expected.}
\end{figure}

Suppose we have $r=10$ replicated studies. For replication $k\in\{1, 2,\ldots, r\}$, (i) the sample size $n_k$ is generated at random from $\{100, 101, \ldots, 200\}$ with replacement, and (ii) a mean $\mu_k$ is generated at random from $\mbox{Uniform}(\mu_{lb}, \mu_{ub})$, for some $\mu_{lb}< \mu_{ub}$. Finally, $n_k$ observations are generated from $\mbox{N}(\mu_k, 1)$ that emulate the data corresponding to the $k$-th replication study. We want to test the hypotheses:
\begin{align}
\mbox{H}_0: \mu\in(-\delta, \delta)\quad \text{against}\quad \mbox{H}_1: \mu\in (-\infty, -\delta]\cup [\delta, \infty).
\end{align}
Please refer to Figure \ref{plot:meta_sim_mom} for results. In the left panel of Figure \ref{plot:meta_sim_mom}, the means $\mu_k,\ k =1,\ldots, 10$ are generated at random from $\mbox{Uniform}(0.0, 0.1)$. Further, if we fix $\delta = 0.1$, the above data-generating scheme can be termed as "data generated under null". The combined Bayes factor function (in red) attains its maxima within the null interval (to the left of the dotted blue line), as expected. In the right panel, the means $\mu_k,\ k =1,\ldots, 10$ are generated at random from $\mbox{Uniform}(0.2, 0.3)$. For varying $\delta$, we present the log Bayes factor for the individual studies, as well as for the meta-analysis. Further, if we fix $\delta = 0.1$, the above data-generating scheme can be termed as "data generated under alternative". The combined Bayes factor function (in red) attains its maxima outside the null interval (to the right of the dotted blue line), as expected.

\subsection{Applications}
\subsubsection{IMAGINE 5 and BIL Asian Population Studies}

\begin{figure}[!htb]
        \center{\includegraphics[height = 5cm, width = 16cm]
        {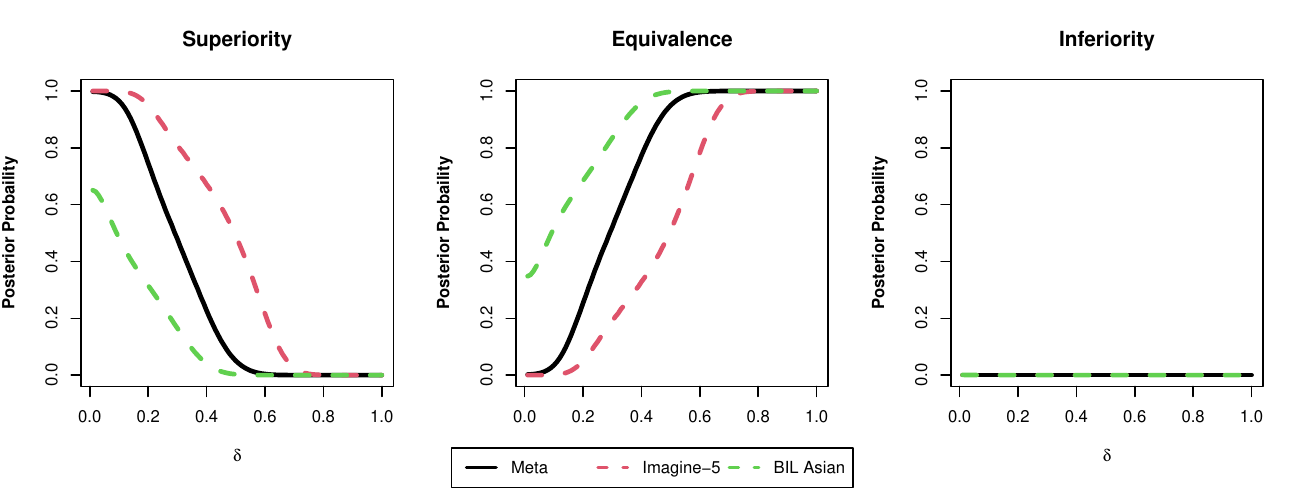}}
        \caption{\label{meta_data_mom}\textbf{ Meta analysis of IMAGINE 5 and BIL Asian population studies (Moment prior)}.  We present the posterior probability of the three competing hypotheses, e.g, inferiority (hypothesis 3), equivalence (hypothesis 2), superiority (hypothesis 1) of BIL compared to Insulin Glargine.}
\end{figure}

To demonstrate the applicability of the proposed meta-analysis approach in practice, along with the IMAGINE 5 study data introduced in sub-section \ref{ssec:bidj}, we utilize the information from the BIL Asian population study as well. BIL Asian population was a Phase 3, open-label, randomized, parallel, 26-week treatment study comparing Basal Insulin Peglispro (BIL) with Insulin Glargine as basal insulin treatment in combination with oral anti-hyperglycemia medications in Asian insulin-naïve patients with type 2 Diabetes Mellitus \citep{biaq2019}. Similar to the IMAGINE 5 study \ref{ssec:bidj}, the primary objective of this study was the same as the IMAGINE 5 study's primary objective, to demonstrate that Insulin Peglispro was noninferior to Insulin Glargine for the change in HbA1c from baseline to 26 weeks of treatment in Asian insulin-naïve patients with T2DM using a noninferiority margin (NIM) of 0.4\%. For the IMAGINE 5 and BIL Asian population studies, if we want to test for superiority, equivalence, and inferiority of BIL compared to Insulin Glargine, we test the competing hypotheses:
\begin{align}
    \mbox{H}_1: \mu_1 - \mu_2 \leq \delta_1 = -0.4,\ \mbox{H}_2: -0.4 = \delta_1 < \mu < \delta_2 = 0.4,\ \mbox{H}_3: \mu \geq \delta_2 = 0.4
\end{align}
respectively, where $\mu_1$ and $\mu_2$ refer to the mean reduction in hemoglobin A1c (HbA1c) from baseline to 26 weeks of treatment in patients with T2DM, with the administration of BIL and Insulin Glargine, respectively. Refer to Figure \ref{meta_data_mom} that shows the posterior probability assigned to the three competing hypotheses for varying values of $\delta$. At $\delta = 0.4\%$, the proposed meta-analysis procedure provides the maximum posterior probability to the hypothesis that BIL is superior to Insulin Glargine with respect to the change in hemoglobin A1c (HbA1c) from baseline at 26 weeks of treatment in patients with T2DM.

\subsubsection{End-to-End Bayesian Testing Framework}
The following application of the proposed Bayes factor-based testing mechanism is in connection with the issue raised in the Bayesian Scientific Working Group (BSWG) Newsletter, July 2023, which states: "Substantial historical data may be available on the active-control and placebo before an active controlled trial is planned in clinical development. Bayesian approaches provide a natural framework for synthesizing the historical data that can effectively be used in designing a non-inferiority clinical trial. Despite a flurry of recent research activities in this area, there are still substantial gaps in the recognition and acceptance of such applications in clinical trial development."

In a shared spirit, for coherent reporting and interpretation of clinical trial outcomes and subsequent decision-making, we present an end-to-end Bayesian hypothesis testing framework, crucially exploiting our proposed meta-analysis framework. In practice, several Phase I, Phase II, and Phase III studies are often conducted to infer the safety and efficacy of a new drug compared to a comparator. For example, in the Basal Insulin Peglispro (BIL) data package, 5 Phase I, 2 Phase II, and 1 Phase III studies were conducted to demonstrate that BIL is non-inferior to standard-of-care insulin for the change in hemoglobin A1c (HbA1c) from baseline to 26 weeks of treatment in patients with T2DM treated with basal insulin alone or in combination with OAM(s) using a noninferiority margin (NIM) of $0.4\%$. In what follows, we describe a seamless Bayesian testing framework to effectively pool information across multiple Phase II and Phase III studies. In the description of our framework, we intentionally leave out the Phase I trials, but it can be readily integrated into the pipeline.

\begin{figure}[!htb]
        \center{\includegraphics[height = 7cm, width = 18cm]
        {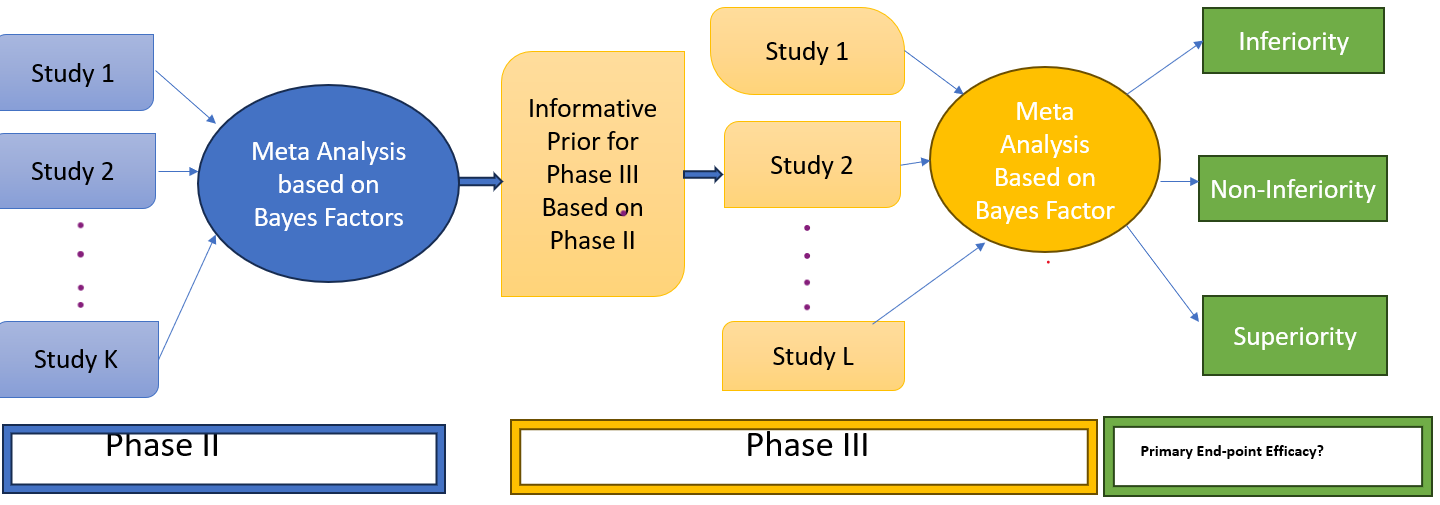}}
        \caption{\label{end_to_end}\textbf{End-to-end Bayesian Testing Pipeline}. }
\end{figure}

Suppose that, in one or more Phase I trials, a new Drug (Drug 1) was compared to an existing Drug (Drug 2), and it met the safety and efficacy endpoints, and the manufacturer decided to move from Phase I to Phase II. In phase II, suppose $K$ independent clinical trials  are conducted to compare Drug 1 to Drug 2 with respect to the primary endpoint. If we want to test for non-inferiority and inferiority of Drug 1 compared to Drug 2, we shall test the competing hypotheses
\begin{align}
    &\mbox{H}_0: \mu_1 -\mu_2< \delta_2\quad \text{(Non-inferiority)}; \quad\text{against}\quad
    \mbox{H}_1: \mu_1 -\mu_2\geq \delta_2\quad \text{(Inferiority)} 
\end{align}
respectively, where $\delta_2$ is fixed based on clinical considerations or FDA regulations, and $\mu_1$ and $\mu_2$ denote the treatment effect on the primary end-point due to the administration of Drug 1 and 2 respectively. We carry out the testing via the meta analysis approach described in equation \ref{eqn:meta_analysis} pooling information from the $K$ independent phase II studies, and calculate the posterior probabilities of the competing hypotheses
\begin{align*}
    \mbox{P}(\mbox{H}_t\mid \text{Phase II Data}),\quad  t = 0, 1\quad \text{for fixed}\quad \delta_2>0.
\end{align*}
On specification of success probability cutoffs, these posterior probabilities facilitate the go/no-go decision to Phase III trials, with respect to the primary endpoint of the study. If the success probability cutoffs are not met, the drug is terminated; otherwise move to Phase III.

In Phase III, suppose $L$ independent clinical trials are conducted to compare Drug 1 to Drug 2 with respect to the primary endpoint. If we again want to test for the superiority, equivalence, or inferiority of Drug 1 compared to Drug 2, we shall again test the multiple competing hypotheses:

\begin{align}
    &\text{H}_0: \mu_1 - \mu_2 < \delta_3\quad \text{(Non-inferiority)}; \quad\text{against}\quad
    \text{H}_1: \mu_1 - \mu_2 \geq \delta_3\quad \text{(Inferiority)} 
\end{align}

Here, $\delta_3$ is fixed based on clinical considerations or FDA regulations, and $\mu_1$ and $\mu_2$ denote the treatment effect on the primary endpoint due to the administration of Drug 1 and 2, respectively. We compute the marginal likelihood of Phase III data:

\begin{align*}
    m(\text{Phase III data}\mid \text{H}_i),\quad i = 0, 1,
\end{align*}

under the competing hypotheses via the meta-analysis approach described in Equation \ref{eqn:meta_analysis}, pooling information from the $L$ independent Phase III studies, and calculate the posterior probabilities:

\begin{align}
    \text{P}(\text{H}_i\mid \text{Phase III Data}) = \frac{m(\text{Phase III data}\mid \text{H}_i)\times \text{P}(\text{H}_i\mid \text{Phase II Data})}{\sum_{i=0}^1 m(\text{Phase III data}\mid \text{H}_i)\times \text{P}(\text{H}_i\mid \text{Phase II Data})},
\end{align}

where $i = 0, 1$ for fixed $\delta_{3} > 0$. On the specification of success probability cutoffs, these posterior probabilities facilitate the go/no-go decision to market, with respect to the primary endpoint of the study. Noteworthy, if Phase II and Phase III data are congruent, i.e., $m(\text{Phase III data}\mid \text{H}_i)$ and $\text{P}(\text{H}_i\mid \text{Phase II Data})$ are in the same order with respect to $i$, the historical borrowing of information across phases of the trial leads to the collection of greater evidence in favor of the hypothesis:

\begin{align*}
    \text{argmax}_{\text{H}_i,\ i = 1, 2, 3} m(\text{Phase III data}\mid \text{H}_i) = \text{argmax}_{\text{H}_i,\ i = 1, 2, 3} \text{P}(\text{H}_i\mid \text{Phase II Data}).
\end{align*}

On the other hand, if Phase II and Phase III data are incongruent, i.e., $m(\text{Phase III data}\mid \text{H}_i)$ and $\text{P}(\text{H}_i\mid \text{Phase II Data})$ are in a different order with respect to $i$, we continue to coherently pool evidence in favor of the hypotheses across Phase II and III. Refer to Figure \ref{end_to_end} for a schematic representation of our proposed end-to-end hypothesis testing framework.

\begin{figure}[!htb]
        \center{\includegraphics[height = 6cm, width = 18cm]
        {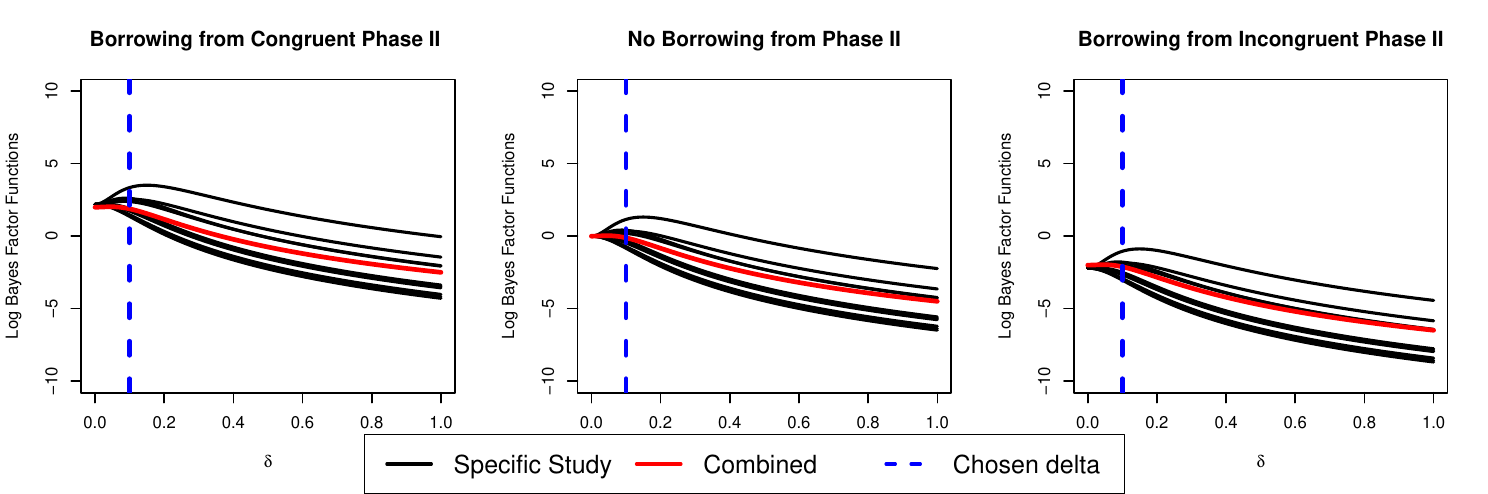}}
        \caption{\label{end_to_end_sim}\textbf{End-to-end Bayesian Testing (Two-way case)}. When the Phase II and Phase III trial outcomes are congruent (incongruent), the evidence in favour of $\mbox{H}_0$ increases (decreases) due to information borrowing across phases. }
\end{figure}

Next, we conduct a small numerical experiment to illustrate the benefits of end-to-end information borrowing in the assessment of clinical trials via the proposed Bayesian testing framework. Suppose, in Phase III, we have $r=10$ replicated studies. For each replication $k\in\{1, 2,\ldots, r\}$:
(i) The sample size $n_k$ is generated from $\{100, 101, \ldots, 200\}$ with replacement.
(ii) A mean $\mu_k$ is generated at random from $\text{Uniform}(0.0, 0.1)$.
Finally, $n_k$ observations are generated from $\text{N}(\mu_k, 1)$ to emulate the data corresponding to the $k$-th replication study. We want to test the hypotheses:

\begin{align}
    \text{H}_0: \mu\in(-\delta, \delta)\quad \text{against}\quad \text{H}_1: \mu\in (-\infty, -\delta]\cup [\delta, \infty),
\end{align}

with $\delta = 0.1$, meaning that the data is generated under $\text{H}_0$. We consider three cases where $\text{P}(\text{H}_0\mid \text{Phase II Data})\in\{0.9, 0.5, 0.1\}$. Case 1 and 3 correspond to situations where the Phase II and Phase III trials are congruent and incongruent, respectively. Case 2 corresponds to no information borrowing. Please refer to Figure \ref{end_to_end_sim} for simulation study results. When the Phase II and III trials are congruent, the evidence in favor of $\text{H}_0$ is increased via information borrowing. On the other hand, when the Phase II and III trials are incongruent, the evidence in favor of $\text{H}_0$ decreases via information borrowing, providing a concrete framework for pooling information across different phases of the clinical trial.

\subsubsection{Competitive Landscape Analysis}

Data driven competitive landscape analysis  is a critical component towards Go/No-Go decision in clinical development. Meta analysis based on Bayes factors can be seamlessly adopted for this task. An example follows. Suppose that, for Type 2 Diabetes populations, we have data available for several competitor drugs available in the market from different Phase 3 insulin trials with reference arm as Insulin Glargine \citep{iope2010,boost2013, element52017,begin3tw2013,beginod2016, instride22022, merck12932018, biamimagine42018,biaq2019,bidjimagine52018}.
For each of the drugs, we test for its superiority, equivalence, and inferiority compared to Insuling Glargine with respect to the primary endpoint of change in HbA1c $\%$ at $24-26$ weeks from baseline,
\begin{align}
    \mbox{H}_{1}: \mu_{com} - \mu_{glar} \in(-\infty, -\delta], \quad
    \mbox{H}_2: \mu_{com} - \mu_{glar}\in(-\delta,\ \delta),\quad
    \mbox{H}_3: \mu_{com} - \mu_{glar}\in [\delta, \infty).
\end{align}
\begin{figure}[!htb]
        \center{\includegraphics[height = 4cm, width = 18cm]
        {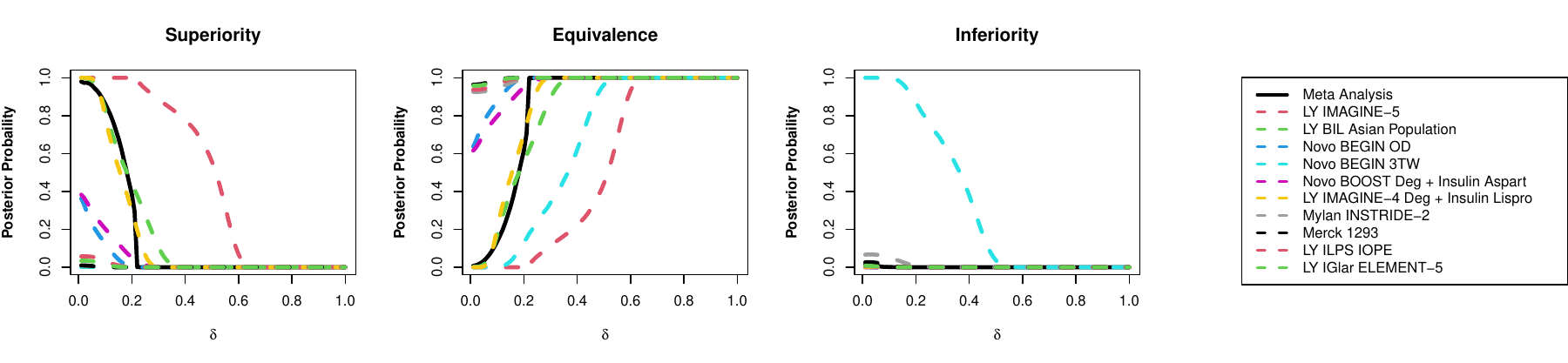}}
        \caption{\label{competitive_landscape_mom}\textbf{Competitive Landscape}. Meta-analysis with outcomes from different Phase 3 insulin trials for T2D  with the reference arm Insulin Glargine with respect to  change in HbA1c $\%$ from baseline at $24/26$ weeks.}
\end{figure}
In Figure \ref{competitive_landscape_mom}, the specific colored lines in the three panels present the posterior probabilities of superiority, equivalence, and inferiority of a specific drug compared to Insulin Glargine, for varying length of the region of clinical equivalence, i.e varying value of $\delta$. The solid black line in the three panels present the posterior probabilities of superiority, equivalence, and inferiority of a hypothetical ``average" drug in the competitive landscape, compared to Insulin Glargine.
This meta-analysis can inform future  insulin trials for Type 2 Diabetes populations with the reference arm Insulin Glargine. In particular, future go/no-go decisions can be based upon efficacy of new potential drug in comparison to existing drugs in the competitive landscape.

\section{Discussion}

The American Statistical Association's statement on statistical significance and P-values highlighted issues with their lack of transparency and inability to measure evidence or effect size. In the context of reporting clinical trial outcomes, our article advocates for the interval null hypothesis framework and Bayes factors based on common t-test statistics to address these challenges. Moreover, to define the Bayes Factors, we adopt non-local priors that are known to facilitate quicker accumulation of evidence in favor of the true  hypothesis, which bears important consequences with regard to the choice of sample size in clinical trials. We applied the proposed approach to several hypothesis testing scenarios in the context of clinical trials, including case studies related to testing for the superiority, equivalence, or inferiority of BIL compared to Insulin Glargine. We also extended the methodology to the meta-analysis setup and conducted real data analysis to demonstrate its practical utility.

Noteworthy future avenues of inquiry include the extension of the proposed frameworks to other hypothesis testing setups (e.g., Analysis of Variance, goodness of fit \citep{10.1214/009053604000000616}, mixture models \citep{rossell2019choosing, chakraborty2023fair, e26010063, chakraborty2023scalable}) in the clinical trial context, and designing corresponding non-local priors. For replicated studies, we may want to develop a meta-analysis framework with dependent non-local priors,  to address the scenario where the studies under consideration cannot be considered independent. Theoretical results on the Bayes factor consistency rates under our setup would require substantial work. While it is well beyond the scope of the current article, it perhaps warrants separate future investigation. On the application front, further numerical experiments and case studies related to other therapeutic areas, including weight loss, heart disease, Atopic dermatitis, etc., would be valuable too.

\section*{ACKNOWLEDGMENT}
We extend our gratitude to Dr. Kofi P. Adragni for his valuable suggestions and guidance, which greatly contributed to the enhancement of the project during Abhisek's summer internship at Eli Lilly and Company. Additionally, we would like to thank Dr. Yongming Yu and Dr. Micheal Sonksen, who read a draft paper and provided insightful comments.
\section*{CONFLICT OF INTEREST STATEMENT}
Majority of work was done when Abhisek was a summer intern at Eli Lilly and Company. 
\section*{DATA AVAILABILITY STATEMENT}
Data are not shared.

\bibliography{references}

\end{document}